\documentclass[preprint,showpacs,preprintnumbers,amsmath,amssymb]{revtex4}

\usepackage{graphicx}
\usepackage{dcolumn}
\usepackage{bm}
\usepackage{color} 
\newcommand{\tcr}{\textcolor{red}}
\newcommand{\tcb}{\textcolor{blue}}

\begin{document}

\preprint{APS/123-QED}

\title{Quantum noise of a Michelson-Sagnac interferometer with translucent mechanical oscillator}

\author{Kazuhiro Yamamoto}
 \email{kazuhiro.yamamoto@aei.mpg.de}
\author{Daniel Friedrich}
\author{Tobias Westphal}
\author{Stefan Go{\ss}ler}
\author{Karsten Danzmann}
\author{Roman Schnabel}
\affiliation{%
Institut f\"{u}r Gravitationsphysik, Leibniz Universit\"{a}t Hannover and Max-Planck-Institut f\"{u}r Gravitationsphysik (Albert-Einstein-Institut),
Callinstrasse 38, D-30167 Hannover, Germany.}%

\author{Kentaro Somiya}
\affiliation{%
Theoretical Astrophysics, California Institute of Technology, Pasadena, California, 91125}%

\author{Stefan L. Danilishin}
\affiliation{%
Physics Faculty, Moscow State University, Moscow 119992, Russia}%

\date{\today}

\begin{abstract}

Quantum fluctuations in the radiation pressure of light can excite
stochastic motions of mechanical oscillators thereby realizing a
linear quantum opto-mechanical coupling. When performing a precise
measurement of the position of an oscillator, this coupling
results in quantum radiation pressure noise. Up to now this effect
has not been observed yet. Generally speaking, the strength of
radiation pressure noise increases when the effective mass of the
oscillator is decreased or when the power of the reflected light
is increased. Recently, extremely light SiN membranes
($\approx$\,100\,ng) with high mechanical Q-values at room
temperature ($\geq 10^6$) have attracted attention as low thermal
noise mechanical oscillators. However, the power reflectance of
these membranes is much lower than unity ($<$ 0.4 at a wavelength
of 1064\,nm) which makes the use of advanced interferometer
recycling techniques to amplify the radiation pressure noise in a
standard Michelson interferometer inefficient. Here, we propose
and theoretically analyze a Michelson-Sagnac interferometer that
includes the membrane as a common end mirror for the Michelson
interferometer part. In this new topology, both, power- and
signal-recycling can be used even if the reflectance of the
membrane is much lower than unity. In particular, signal-recycling
is a useful tool because it does not involve a power increase at
the membrane. We derive the formulas for the quantum radiation
pressure noise and the shot noise of an oscillator position
measurement and compare them with theoretical models of the
thermal noise of a SiN membrane with a fundamental resonant
frequency of 75\,kHz and an effective mass of 125\,ng. We find
that quantum radiation pressure noise should be observable with a
power of 1\,W at the central beam splitter of the interferometer
and a membrane temperature of 1\,K.
\end{abstract}



\maketitle

\section{Introduction}

Laser interferometers are among the most sensitive measurement
devices ever built. The interferometric gravitational wave
detectors in their first generation achieve a linear noise
spectral density for the displacement measurement of as low as
$10^{-19}$\,m/Hz$^{1/2}$ \cite{Stan}. The gravitational wave
detectors of the second generation \cite{AdLIGO,AdVIRGO,LCGT} are
designed to have a ten times better sensitivity. The sensitivity
of these interferometers will be limited by quantum radiation
pressure noise \cite{Caves} at low audio-band Fourier frequencies
and by photon shot noise \cite{Caves} at higher frequencies. While
the shot-noise limited regime of laser interferometers has been
fully investigated and even interferometers with squeezed shot
noise were demonstrated \cite{XWK87, GSYL87,KSMBL02,VCHFDS05}, the
radiation pressure noise has not yet been observed. The
experimental investigation of this quantum measurement regime is
crucial in view of future gravitational wave detectors. It is also
interesting from the fundamental physics point of view because the
successful observation of the quantum radiation pressure noise
will confirm the seminal principle of back-action in a
(continuous) quantum measurement \cite{SQL,BKh96} when one quantum
system (the light) serves as a probe for another quantum system
(the mechanical oscillator).

In order to observe quantum radiation pressure noise, the
mechanical oscillator under investigation should have a high
susceptibility to radiation pressure, i.e.\ a low effective mass
\cite{Kippenberg}. The motion of the oscillator should show a low
thermal noise,
i.e.\ the oscillator should have a high mechanical Q-value.
Recently, commercially available SiN membranes have attracted a
lot of attention and were considered for experiments aiming for
nonlinear quantum effects of mechanical oscillators, i.e.\ the
observation of quantum jumps \cite{Harris1,Harris2,Harris3}. The
membranes typically have an effective mass of the order of
100\,ng, a thickness of about 100\,nm and a surface area of about
1\,mm$^2$. The Q-values at their fundamental resonant frequency at
about 100\,kHz were measured to 10$^6$ at room temperature and
10$^7$ at 300\,mK \cite{Harris2}. These properties of the
membranes make them also interesting for experiments aiming for
the linear quantum regime of light/matter systems as considered
here. In such experiments the dynamic (small) displacement of the
oscillator scales linearly with the amplitude of the incident
light amplitude modulation, and the amplitude of the reflected
light phase modulation scales linearly with the (small)
displacement of the oscillator. In an actual experiment two
identical oscillators might be used as the end mirrors of a simple
Michelson interferometer with homodyne readout. A problem arises
due to the rather low power reflectance of the membrane
($<\,40\,\%$ at a wavelength of 1064\,nm) because the
interferometer techniques of power-recycling and signal-recycling
\cite{Drever1, Drever2, Meers} to amplify the radiation pressure
noise are not efficient for translucent mechanical oscillators.
Power-recycling is used to resonantly enhance the light power
inside the interferometer without reducing the signal bandwidth of
the interferometer. Signal-recycling is used to resonantly enhance
the signal without increasing the laser power inside the
interferometer.

\begin{figure}
\begin{minipage}{12cm}
\includegraphics[width=12cm]{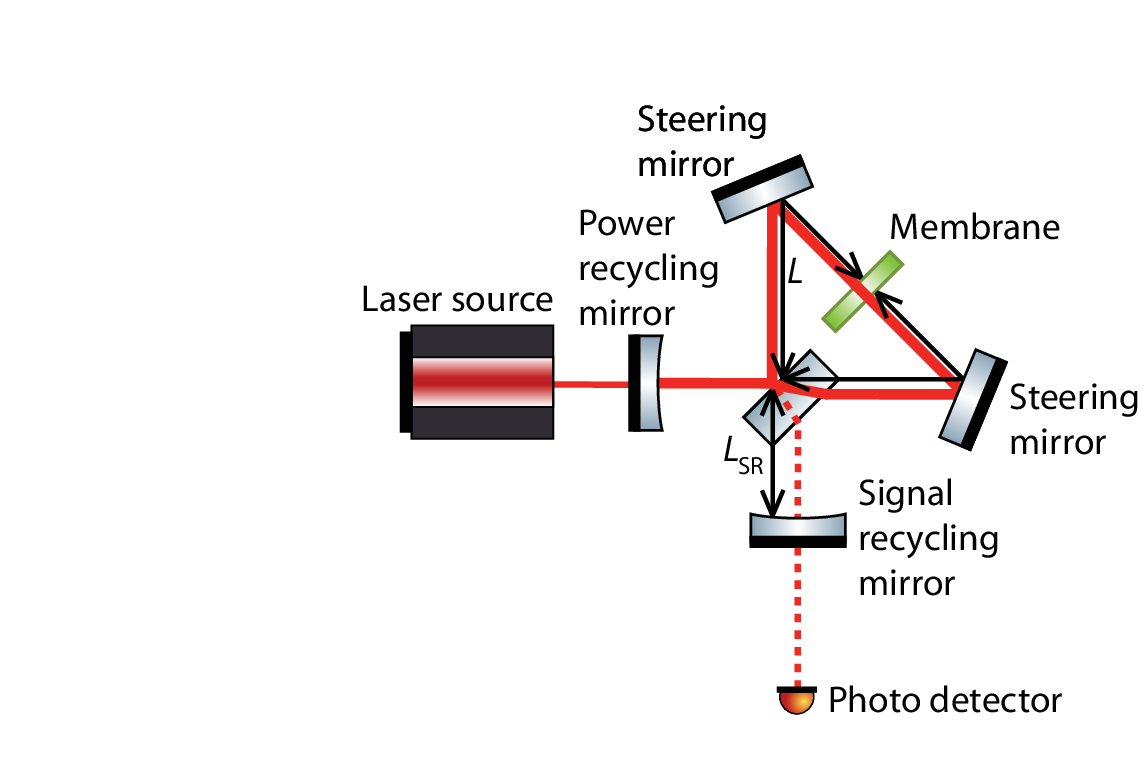}
\end{minipage}
\caption{\label{MichelsonSagnac}Schematic of a power- and
signal-recycling Michelson-Sagnac interferometer. The translucent
mechanical oscillator forms the joint end mirror of the two
Michelson arms. While the displacement of the oscillator is
measured via the Michelson mode, the transmitted light is stored
in the Sagnac mode. The combination of Michelson and Sagnac mode
enables the resonant enhancement of both, the laser power and the
displacement signal by placing power- and signal-recycling mirrors
in the input and output ports of the interferometer, respectively.
The parameter $L$ is the length of the Michelson interferometer
arms, and $L_{\rm SR}$ is the distance between the beam splitter
and the signal-recycling mirror.}
\end{figure}

Here we propose a recycling Michelson-Sagnac interferometer in
order to access the radiation pressure noise regime of a
translucent mechanical oscillator. In this interferometer topology
a single oscillator forms the joint end mirror of a Michelson
interferometer with folded arms \cite{Corbitt}. While the
reflected light forms a Michelson mode, the light transmitted
through the oscillator forms a Sagnac mode (see
Fig.~\ref{MichelsonSagnac}). Since all the light, including the
signal, is kept within the two modes, power- as well as
signal-recycling can be used to enhance the radiation pressure
noise. We calculate the quantum noise of the Michelson-Sagnac
interferometer and discuss the role of the power- and
signal-recycling techniques for the observation of quantum
radiation pressure noise. We finally compare the power spectral
densities of radiation pressure noise, shot noise, and thermal
noise for an oscillator displacement measurement and answer the
question as to what values of light power and temperature of the
membrane would allow an observation of radiation pressure noise.

\section{Quantum noise of a Michelson-Sagnac interferometer}

For a displacement measurement of a mechanical oscillator, photon
shot noise as well as radiation pressure noise has to be taken
into account. In a Michelson-Sagnac interferometer the
corresponding spectral densities differ from those in a simple
Michelson or Sagnac \cite{Chen2,Danilishin,Chen3} interferometer
due to the interference of the Michelson and the Sagnac modes.


\subsection{Photon shot noise of membrane displacement measurement}

Let us first consider a bare laser Sagnac interferometer what does
not rotate. If the beam splitter has a perfect 50\%/50\% splitting
ratio, all the incident light is back reflected towards the laser
source. It is assumed that the Sagnac mode has a waist at half the
round trip length. In this case a reflecting plane surface with
amplitude reflectance $r$ (a membrane) can be put into the waist
in such a way that the light is not scattered out of the
interferometer and additionally creates a Michelson mode that is
sensitive to the motion of the membrane. Note that the length
change in each Michelson arm is twice the displacement of the
membrane, and the differential length change in the two arms is
twice times that change. Then the power at the output port is
given by
\begin{eqnarray}
P_{\rm out}&=&\frac{r^2 P_0}{2}\left[1-\cos\left(\Phi_0 +
\frac{8\pi}{\lambda}x \right)\right], \label{shot noise first}
\end{eqnarray}
where $x$ is the displacement of the membrane from its operating
point $\Phi_0$ and $P_0$ is the incident power. If the
displacement is much smaller than the wavelength of light
$\lambda$, Eq.~(\ref{shot noise first}) can be approximated using
a Taylor expansion yielding
\begin{eqnarray}
P_{\rm out}&\sim& \frac{r^2
P_0}{2}\left(1-\cos\Phi_0+\frac{8\pi}{\lambda}x\sin\Phi_0\right).
\end{eqnarray}
The single-sided linear spectral density $\sqrt{G_{\rm out}}$ of
the shot noise of the light at the output port is described as
\cite{Edelstein}
\begin{equation}
\sqrt{G_{\rm out}}=\sqrt{2 \hbar \omega_0 P_{\rm out}} =
\sqrt{\frac{4 \pi \hbar c P_{\rm out}}{\lambda}} \label{Vacuum},
\end{equation}
where $\hbar$ is the reduced Planck constant, $\omega_0~(=2\pi
c/\lambda)$ is the angular frequency of light, and $c$ is the
speed of light. The signal-normalized shot noise is then given by
\begin{eqnarray}
\sqrt{G_{\rm shot}} &=& \sqrt{\frac{4 \pi \hbar c P_{\rm
out}}{\lambda}}\left[\left|\frac{\partial P_{\rm out}}{\partial
x}\right|_{x=0}\right]^{-1} = \sqrt{\frac{2 \pi \hbar c r^2 P_0
(1-\cos\Phi_0)}{\lambda}}\,\frac{2}{r^2 P_0}\,\frac{\lambda}{8\pi
|\sin \Phi_0|}
\nonumber\\
&=&\sqrt{\frac{\hbar c \lambda}{16 \pi r^2
P_0}}\,\frac1{|\cos(\Phi_0/2)|}. \label{shotpre}
\end{eqnarray}
Hence, for $\Phi_0=0$ corresponding to the dark fringe, the
signal-normalized shot noise is minimum for given a power $P_0$
\begin{equation}
\sqrt{G_{\rm shot}}=\sqrt{\frac{\hbar c \lambda}{16 \pi r^2 P_0}}.
\label{shot}
\end{equation}

\subsection{Quantum radiation pressure noise of membrane displacement measurement}

\begin{figure}
\begin{minipage}{8.6cm}
\includegraphics[width=4cm]{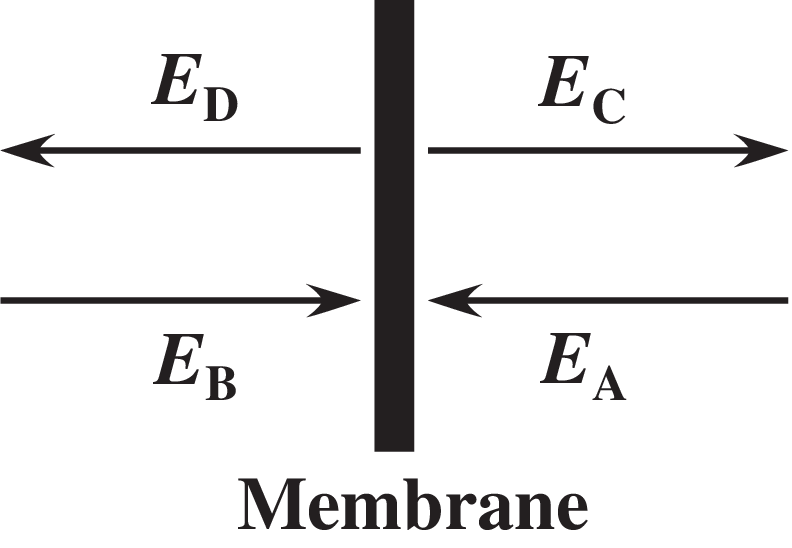}
\end{minipage}
\caption{\label{membrane}
The four arrows represent the incident light fields ($E_{\rm A}, E_{\rm B}$)
and outgoing interference of the reflected and transmitted light
($E_{\rm C}, E_{\rm D}$). These four fields give
rise to radiation pressure effects at the membrane.
}
\end{figure}
The radiation pressure force is equal to the momentum transferred
from the light to the membrane per unit time. This is the
difference of the light field momentum flux between outgoing and
incident beam, at the reflecting surface integrated over the
effective area ${\cal A}$ of the light beam. The total momentum
flux results from two pairs of incident ($E_{\rm A},E_{\rm B}$)
and outgoing ($E_{\rm C}, E_{\rm D}$) travelling waves (see
Fig.~\ref{membrane}). The momentum density for each wave is equal
to $\epsilon_0 \overline{|E_{\alpha}|^2}/c$, where $\alpha={\rm
A,B,C,D}$. $E_{\alpha}$ is the electric field strain in the
corresponding wave, $\epsilon_0$ is the permittivity of vacuum,
and $\overline{|E_{\alpha}|^2}$ stands for time average of
$|E_{\alpha}|^2$ over many periods of light oscillation. Taking
into account the directions of light propagation and choosing
signs in front of each wave momentum flux correspondingly, one
gets the following expression for the radiation pressure force
exerted by light on the membrane
\begin{equation}
F_{\rm RP} = c \times {\cal A}\frac{\epsilon_0}{c}
\left[\left(\overline{|E_{\rm C}|^2} -\overline{|E_{\rm
D}|^2}\right)-\left(\overline{|E_{\rm B}|^2} - \overline{|E_{\rm
A}|^2} \right) \right]. \label{radiation pressure noise force}
\end{equation}
DC-components of the force in Eq.~(\ref{radiation pressure noise
force}) cancel because of the symmetry of the membrane. However,
vacuum fluctuations enter the interferometer from the output port.
This results in quantum amplitude fluctuations via the
interference with the carrier light. In this case, there is a
perfectly negative correlation between the fluctuations of $E_{\rm
A}$ and $E_{\rm B}$ because of the energy conservation at the beam
splitter \cite{Caves}. This is the origin of quantum radiation
pressure noise. The vacuum fluctuations from the input port do not
lead to a displacement of the membrane because the correlation
between $E_{\rm A}$ and $E_{\rm B}$ is perfectly positive.

The incident light field amplitudes are (In Ref.~\cite{Kimble},
this formula is written in cgs Gauss units. However, in this
paper, we use SI units.)
\begin{eqnarray}
E_{\rm A} &=& \frac1{\sqrt{2}}\sqrt{\frac{\hbar \omega_0}{{\cal
A}c\epsilon_0}}\left[\sqrt{2}{\cal D}+{\cal E}_{\rm v1}\right]\cos(\omega_0 t) +
\frac1{\sqrt{2}} \sqrt{\frac{\hbar \omega_0}{{\cal
A}c\epsilon_0}}{\cal E}_{\rm v2} \sin (\omega_0
t),\label{EA}\\
E_{\rm B} &=& \frac1{\sqrt{2}}\sqrt{\frac{\hbar \omega_0}{{\cal
A}c\epsilon_0}}\left[\sqrt{2}{\cal D}-{\cal E}_{\rm v1}\right]\cos(\omega_0 t) -
\frac1{\sqrt{2}} \sqrt{\frac{\hbar \omega_0}{{\cal
A}c\epsilon_0}}{\cal E}_{\rm v2} \sin (\omega_0
t).\label{EB}
\end{eqnarray}
Here $\cal D$ represents the amplitude of the carrier before the
beam splitter, in such a way that ${\cal D}^2$ is the number of
photons per unit time in the beam. The relation between ${\cal D}$
and the incident power $P_0$ is given by
\begin{equation}
P_0
= \hbar \omega_0 {\cal D}^2. \label{intensity}
\end{equation}
The parameters ${\cal E}_{\rm v1}$ and ${\cal E}_{\rm v2}$ stand
for amplitude and phase quadrature of vacuum fields, respectively,
that enter the interferometer from the output port and propagate
to the membrane.
The phase shift of transmittance of the membrane is not
independent of that of the reflectance owing to energy
conservation ($\overline{E_{\rm A}^2}+\overline{E_{\rm
B}^2}=\overline{E_{\rm C}^2}+\overline{E_{\rm D}^2}$). Since the
membrane is symmetric, the phase difference between the
reflectance and transmittance must be $\pi/2$. It is possible to
assume a phase shift in the reflection of $\pi/2$, while the phase
shift in transmission is zero, without losing generality.
The outgoing fields are written as
\begin{eqnarray}
E_{\rm C} &=& \frac1{\sqrt{2}}\sqrt{\frac{\hbar \omega_0}{{\cal
A}c\epsilon_0}}\left[\sqrt{2}t{\cal D}-t{\cal E}_{\rm v1}+r{\cal
E}_{\rm
v2}\right]\cos(\omega_0 t)\nonumber\\
&& + \frac1{\sqrt{2}} \sqrt{\frac{\hbar \omega_0}{{\cal
A}c\epsilon_0}}\left[-\sqrt{2}r{\cal D}-r{\cal E}_{\rm v1}-t{\cal
E}_{\rm v2}\right]\sin (\omega_0
t),\label{EC}\\
E_{\rm D} &=& \frac1{\sqrt{2}}\sqrt{\frac{\hbar \omega_0}{{\cal
A}c\epsilon_0}}\left[\sqrt{2}t{\cal D}+t{\cal E}_{\rm v1}-r{\cal
E}_{\rm
v2}\right]\cos(\omega_0 t)\nonumber\\
&& + \frac1{\sqrt{2}} \sqrt{\frac{\hbar \omega_0}{{\cal
A}c\epsilon_0}}\left[-\sqrt{2}r{\cal D}+r{\cal E}_{\rm v1}+t{\cal
E}_{\rm v2}\right]\sin (\omega_0 t).\label{ED}
\end{eqnarray}
The parameters $r$ and $t$ are amplitude reflectance and
transmittance of the membrane, respectively.

Substituting Eqs.~(\ref{EA}), (\ref{EB}), (\ref{EC}), and
(\ref{ED}) in Eq.~(\ref{radiation pressure noise force}), we
obtain the force exerted by the radiation on the membrane
\begin{equation}
F_{\rm RP} = \frac{2}{c} \sqrt{2 \hbar \omega_0 P_0}\,r^2 {\cal
E}_{\rm v1} + \frac{2}{c} \sqrt{2 \hbar \omega_0 P_0}\,rt {\cal
E}_{\rm v2}. \label{radiation pressure noise force 2}
\end{equation}
Let us assume that the field entering from the output port is a
vacuum state. The single-sided spectral densities of ${\cal
E}_{v1}$ and ${\cal E}_{v2}$ are equal to 1 (this spectral density
has no dimension because the dimension of ${\cal E}_{\rm v1}$ and
${\cal E}_{\rm v2}$ are the same as that of ${\cal D}$, which is
the square root of photon number per unit time) while their
cross-correlation vanishes \cite{Kimble}, thus one can easily get
the following expression for single-sided spectral density of
radiation pressure force
\begin{equation}
\sqrt{G_{F_{\rm RP}}} = \sqrt{\left(\frac{2}{c} \sqrt{2 \hbar
\omega_0 P_0}\,r^2\right)^2+\left(\frac{2}{c} \sqrt{2 \hbar
\omega_0 P_0}\,rt \right)^2}= \sqrt{\frac{16 \pi \hbar r^2 P_0}{c
\lambda}}. \label{radiation pressure noise force 3}
\end{equation}
The motion of the membrane caused by this force is
\begin{eqnarray}
\sqrt{G_{\rm rad}}&=& H \sqrt{\frac{16 \pi \hbar r^2 P_{0}}{c\lambda}},\label{rad}\\
H &=&\left|\frac1{-m_{\rm mem}\,(2 \pi f)^2+m_{\rm mem}\,(2 \pi
f_{\rm mem})^2(1+{\rm i}f/(Qf_{\rm mem}))}\right|\label{Hmem}.
\end{eqnarray}
The function $H$ shows the (complex) mechanical susceptibility of
the membrane, defined by its effective mass $m_{\rm mem}$
\cite{Yamamotomode}, resonant frequency $f_{\rm mem}$, and
mechanical Q-value $Q_{\rm mem}$.

\subsection{Power- and signal-recycling techniques}\label{PRSR}

The membrane in the Michelson-Sagnac interferometer can be
positioned such that not only the Sagnac mode but also the
Michelson mode is on a dark fringe at the signal port. In this
case (almost) all the input light power is back-reflected towards
the laser source and power- and signal-recycling \cite{Drever1,
Drever2, Meers} can be used in order increase the quantum
radiation pressure noise. Power- and signal-recycling techniques
are realized via additional mirrors (see
Fig.~\ref{MichelsonSagnac}) which together with the
Michelson-Sagnac interferometer form cavities for carrier light
(power-recycling) and signals and vacuum fluctuation entering the
signal port (signal-recycling). In this paper, we consider only
the signal-recycling cavity tuned to the carrier frequency
\cite{tune definition}.

In the case of power-recycling \cite{Drever1, Drever2}, the
recycling cavity enhances the incident (carrier) power by a factor
$g_{\rm PR}$, which is the power-recycling (energy) gain. The
incident power $P_0$ in formulae of shot noise Eq.~(\ref{shot})
and radiation pressure noise Eq.~(\ref{rad}) is replaced by
$g_{\rm PR}P_0$.
In case of using signal-recycling \cite{Meers}, the amplitude of
sidebands caused by the membrane motion ($|\partial P_{\rm
out}/\partial x|$ in Eq.~(\ref{shotpre})) and the vacuum
fluctuations from the output port are amplified by $\sqrt{g_{\rm
SR}}$, which is the signal-recycling (amplitude) gain.

It should be noted that the sideband frequency of the signal and
the corresponding vacuum fluctuation is preferred to be smaller
than the signal-recycling cavity bandwidth $f_{\rm SR}$. If the
sideband frequency is larger than $f_{\rm SR}$, amplification of
signal sidebands and vacuum fluctuation decreases. Hence, the
signal-normalized shot noise increases, while the radiation
pressure noise decreases. If the lengths of the two optical paths
from the beam splitter to membrane are equal, the cut-off
frequency $f_{\rm SR}$ is inversely proportional to the summation
of the distance between beam splitter and signal-recycling mirror
$L_{\rm SR}$ and the arm length $L$ (see
Fig.~\ref{MichelsonSagnac}). The formulae of shot noise and
radiation pressure noise with power- and signal-recycling are
written as \cite{Meers,Fritschel,Strain}
\begin{eqnarray}
\sqrt{G_{\rm shot}}&=& \sqrt{\frac{\hbar c \lambda}{16 \pi g_{\rm
PR}\,g_{\rm SR}\,r^2 P_0}} \sqrt{1+\left(\frac{f}{f_{\rm
SR}}\right)^2},\label{shot-noise SR}\\
\sqrt{G_{\rm rad}}&=&H \sqrt{\frac{16 \pi \hbar g_{\rm PR}\,g_{\rm
SR}\,r^2 P_{0}}{c\lambda}} \frac1{\sqrt{1+\left(f/f_{\rm
SR}\right)^2}},\label{radiation pressure noise SR}\\
{\rm with}\ f_{\rm SR} &=& \frac{c(1-r_{\rm SR})}{4 \pi (L_{\rm
SR}+L)}.\label{cut-off}
\end{eqnarray}
If the reflectance of the recycling mirrors is close unity, but
still lower than the reflectance of the Michelson-Sagnac
interferometer, the recycling gains are
\begin{eqnarray}
g_{\rm PR} &=& \frac{1+r_{\rm PR}}{1-r_{\rm PR}},\label{power gain}\\
g_{\rm SR} &=& \frac{1+r_{\rm SR}}{1-r_{\rm SR}},\label{signal
gain}
\end{eqnarray}
where $r_{\rm PR}$ and $r_{\rm SR}$ are amplitude reflectance of
power- and signal-recycling mirrors, respectively.

\section{Specifications for a linear quantum opto-mechanical coupling}

\begin{table}
\caption{\label{spec} Example specifications of a Michelson-Sagnac
interferometer \cite{Harris2}.}
\begin{ruledtabular}
\begin{tabular}{cc}
Light wavelength ($\lambda$) & 1064\,nm \\
Arm length of Michelson interferometer ($L$) & 0.6\,m \\
Length between signal-recycling mirror and beam splitter ($L_{\rm SR}$) & 3\,cm\\
Amplitude reflectance of signal-recycling mirror ($r_{\rm SR}$) & 0.998 \\
Signal-recycling (amplitude) gain ($\sqrt{g_{\rm SR}}$) & 32 \\
Power at beam splitter ($g_{\rm PR}P_0$)(with/without signal-recycling) & 1\,W/1\,kW \\
Power reflectance of membrane ($r^2$) & 0.35 \\
Resonant frequency of membrane ($f_{\rm mem}$) & 75\,kHz \\
Effective mass of membrane ($m_{\rm mem}$) & 125\,ng \\
Q-value of membrane ($Q_{\rm mem}$) & $10^7$ \\
Temperature of membrane ($T_{\rm mem}$) & 1\,K
\end{tabular}
\end{ruledtabular}
\end{table}

\begin{figure}
\begin{minipage}{8.6cm}
\includegraphics[width=8.6cm]{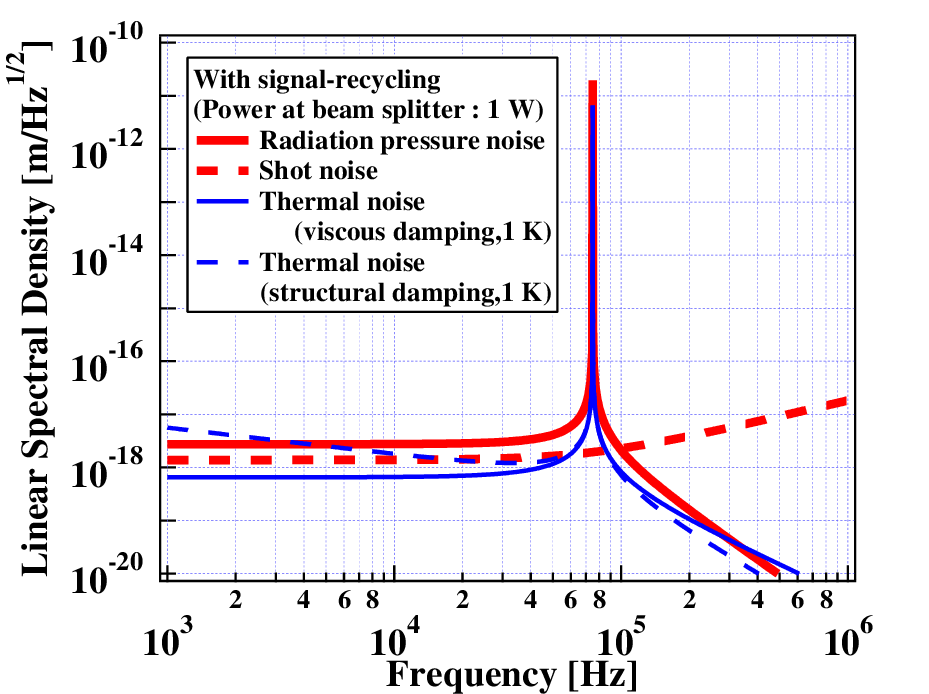}
\end{minipage}
\quad
\begin{minipage}{8.6cm}
\includegraphics[width=8.6cm]{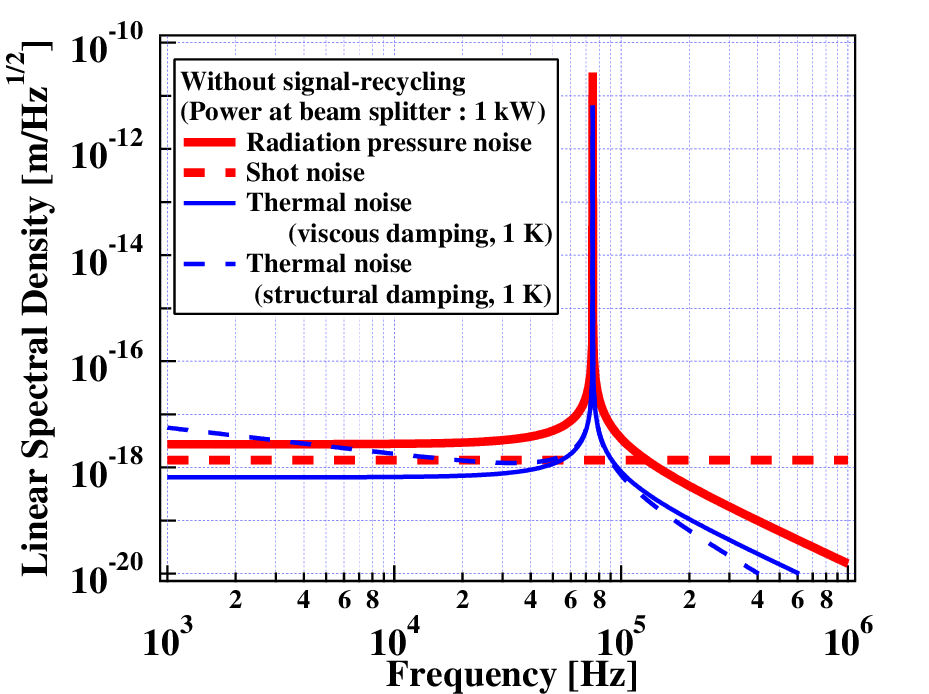}
\end{minipage}
\caption{\label{Sensitivity}Goal sensitivity of the
Michelson-Sagnac interferometer to measure radiation pressure
noise based on the parameters in Table~\ref{spec}. The graphs on
the left and right hand sides show the sensitivity with and
without signal-recycling, respectively. Thick solid and dashed
lines (red in online) are the radiation pressure noise and shot
noise, respectively. Thin solid and dashed lines (blue in online)
are thermal noise at 1\,K in the cases of the viscous damping and
structural damping \cite{Saulson}.}
\end{figure}

In this section we specify a design example of a Michelson-Sagnac
interferometer that should allow for the generation and
observation of the linear quantum opto-mechanical coupling of a
light field with the motion of a SiN membrane. The design
parameters are chosen such that the regime dominated by radiation
pressure noise
can be reached, for some low temperature of the membrane. In our
analysis we in particular compare designs with and without
signal-recycling and found that the implementation of the
signal-recycling technique is beneficial for the purpose of
observing radiation pressure noise. The results of our comparison
are based on the parameters given in Table~\ref{spec}
\cite{Harris2} and are illustrated in Fig.~\ref{Sensitivity}. In
our design example the radiation pressure noise is two times
larger than the shot noise below resonant frequency, and even the
standard quantum limit \cite{SQL,BKh96,Kimble} where
shot noise is equal to radiation pressure noise 
can be reached.
Indeed, as one can see from Eq.~(\ref{radiation pressure noise
SR}), the optical power that determines the level of radiation
pressure noise as well as shot noise always enters the formulae in
the combination: $g_{\rm PR}\,g_{\rm SR}\,P_0/(1+(f/f_{\rm
SR})^2)$. Therefore for frequencies below the signal-recycling
cavity cut-off frequency $f_{\rm SR}$ (76 kHz for our choice of
parameters), one needs $g_{\rm SR}$ times lower optical power
$g_{\rm PR}\,P_0$ at the beam splitter to provide the same level
of radiation pressure noise compared to the case without
signal-recycling. In our example, just 1 W is required when the
signal-recycling technique is adopted compared to 1 kW without
signal-recycling (We need high signal-recycling gain $\sqrt{g_{\rm
SR}}=32$. According to Eq.~(\ref{signal gain}), the loss in the
Michelson-Sagnac interferometer must be smaller than 0.4\%, which
is a realistic value \cite{Schnier}). Signal-recycling reduces the
heat from the absorption in the membrane and beam splitter without
sacrificing the radiation pressure noise.
%
For frequencies $> f_{\rm SR}$, signal-recycling is less effective
as shown in Fig.~\ref{Sensitivity} and Sec.~\ref{PRSR}. The
signal-recycling cut-off frequency should therefore be designed to
be above the observation band where the measurement is performed.

The observation of a quantum opto-mechanical coupling requires a
low enough thermal noise. The off-resonant spectrum of thermal
noise of an oscillator is not precisely known. Here, we consider
two different dissipation mechanisms, namely viscous and
structural damping \cite{Saulson}. The thermal noise formulae are
\begin{eqnarray}
G_{\rm thermal} &=& |H(f)|^2\,\frac{4 k_{\rm B} T_{\rm mem} m_{\rm mem}\,(2\pi f_{\rm mem})}{Q_{\rm mem}}\ ({\rm viscous\ damping}),\label{viscous}\\
G_{\rm thermal} &=& |H(f)|^2\,\frac{4 k_{\rm B} T_{\rm mem} m_{\rm
mem}\,(2\pi f_{\rm mem})^2}{Q_{\rm mem}\,(2 \pi f)}\ ({\rm
structural\ damping}),\label{structure}
\end{eqnarray}
where $k_{\rm B}$ is the Boltzmann constant and $T_{\rm mem}$ is
the temperature of the membrane. Figure \ref{Sensitivity} shows
the thermal noise of the membrane position measurement for the two
dissipation mechanisms. For viscous damping the thermal noise
spectrum is flat below the resonance (solid blue line). For
structural damping the thermal noise spectrum falls with
$1/f^{1/2}$ (dashed blue line) because Fig.~\ref{Sensitivity}
shows the square root of Eqs.~(\ref{viscous}) and
(\ref{structure}). Both thermal noise spectra are plotted for a
membrane temperature of 1\,K. At this temperature the thermal
noise level is about a factor of 3 smaller than the radiation
pressure noise around the membrane resonance frequency. At room
temperature the thermal noise is about 50 times larger, also
partly because the Q-value of the membrane is 10 times smaller
than at about 1\,K \cite{Harris2}. In this case, the power at the
beam splitter must be 3\,kW even if signal-recycling is adopted.
Such a high power can in principle be achieved with the help of
power-recycling. However, light absorption in the membrane and in
the interferometer will most likely cause thermal problems at
these high powers.

In order to reduce the heat, the membrane should be placed at a
node of the standing wave of the Sagnac mode. If the membrane is
at the node, the output port is dark. The recycling techniques are
compatible. Even at the node, the radiation pressure noise is not
zero because the radiation pressure noise depends on the momentum
flux, not energy density as shown in Eq.~(\ref{radiation pressure
noise force}). The signal sideband of the Michelson mode depends
always linearly on the membrane displacement.

The displacement signal of the interferometer can be detected
using a homodyne detector that is realized as a single photo diode
if the Michelson-Sagnac interferometer is operated close to, but
not exactly at a dark fringe. In this case the interferometer
laser field provides the optical local oscillator for homodyne
detection. Alternatively, a \textit{balanced} homodyne detector
can be used if the interferometer is operated exactly at a dark
fringe. Such a detection scheme uses a beam splitter, an external
local oscillator field and two photo diodes and was previously
used as an interferometer read-out, for example in
Refs.~\cite{KSMBL02,VCHFDS05}. A balanced homodyne detector might
be useful in order to increase the reflectance of the
Michelson-Sagnac interferometer allowing for high signal-recycling
gains.

\section{Summary and conclusion}
Interferometer recycling techniques are useful in experiments that
aim for the observation of radiation pressure noise of the
position/displacement measurement of a mechanical oscillator. The
Michelson-Sagnac interferometer topology, as proposed here, is
compatible with power- and signal-recycling, and is able to
incorporate a mechanical oscillator that transmits a major part of
the incident laser light. We have presented spectral densities for
the shot noise and radiation pressure noise for a position
measurement of a translucent SiN membrane. The expressions differ
from those of a usual Michelson interferometer because of the
interference between the beams reflected and transmitted by the
membrane. Signal-recycling can reduce the power in the
interferometer required for pushing the radiation pressure noise
above the shot noise. We have found that the radiation pressure
noise of the interferometer signal is twice as large as the shot
noise below the membrane resonant frequency of 75\,kHz for a laser
power of 1\,W incident on the beam splitter and a signal-recycling
amplitude gain of 32. In order to realize such a high
signal-recycling gain, the Michelson-Sagnac interferometer has to
show a high reflectance. Such a high value is realistic if the
optical loss inside the interferometer is kept to a minimum (less
than 0.4\%) and the interferometer is operated very close to a
dark fringe. For an operation at exactly the dark fringe a
balanced homodyne detector has to be used. The membrane can be
positioned in a node of the standing wave of the Sagnac mode in
order to make absorption smaller without losing the
proportionality between signal field and membrane displacement. If
the membrane temperature is 1\,K, the calculated thermal noise of
the oscillator fundamental mode is below the radiation pressure
noise. This might turn out to be important in order to reduce the
heating of the membrane due to absorbed laser power.

%
%

\begin{acknowledgments}
This work is supported by the Deutsche Forschungsgemeinschaft and
is part of Sonderforschungsbereich 407. K.\,S. is supported by the
Japanese Society for the Promotion of Science. We are grateful to
Yanbei Chen, Thomas Corbitt, and Albrecht R\"{u}diger for useful
comments.

\end{acknowledgments}


\begin{thebibliography}{99}
\bibitem{tune definition}Due to nonzero transmittance,
the membrane modifies the optical path for both Sagnac and
Michelson modes yielding into additional phase shift $\psi = \pm
\arccos r$ between the ingoing and outgoing waves at the beam
splitter. Therefore, the term ``tuned recycling cavity" implies
that the distance betwen the recycling mirror and the beam
splitter should satisfy:
$$L_{\rm PR/SR}^{\rm tuned} = \frac{c}{\omega_0}\Bigl(\pi N + \frac\psi2\Bigr)\,,$$ with $N$ being integer.
\end{thebibliography}

\end{document}